\documentclass[journal,comsoc]{IEEEtran}

\ifCLASSINFOpdf
\else
\fi
\ifCLASSINFOpdf
   \usepackage[pdftex]{graphicx}
\else
\fi
\usepackage{pbox}
\usepackage{stfloats}
\usepackage{epstopdf}
\usepackage{algorithm}
\usepackage{algpseudocode}
\usepackage[table]{xcolor}
\usepackage{pbox}
\usepackage{stfloats}
\usepackage{epstopdf}
\usepackage{color}
\usepackage{amssymb}
\usepackage{graphicx}
\usepackage{pgfplots}
\usepackage{filecontents}
\usepackage{pgfplotstable}
\usepackage{subfig}
\usepackage{url}
\usetikzlibrary{spy}
\usepackage{amsmath}
\usepackage{cite}
\usepackage{flushend}
\usepackage{tabularx}
\usepackage[autostyle]{csquotes}
\usepackage{cleveref}
\usetikzlibrary{patterns}
\usepackage{multirow}
\usepackage{pgf-pie}
\usepackage[utf8]{inputenc}
\usepackage{rotating}
\newcommand*\rot{\rotatebox{90}}
\usepackage{filecontents}

\pgfplotsset{
    box plot/.style={
        /pgfplots/.cd,
        black,
        only marks,
        mark=-,
        mark size=1em,
        /pgfplots/error bars/.cd,
        y dir=plus,
        y explicit,
    },
    box plot box/.style={
        /pgfplots/error bars/draw error bar/.code 2 args={%
            \draw  ##1 -- ++(1em,0pt) |- ##2 -- ++(-1em,0pt) |- ##1 -- cycle;
        },
        /pgfplots/table/.cd,
        y index=2,
        y error expr={\thisrowno{3}-\thisrowno{2}},
        /pgfplots/box plot
    },
    box plot top whisker/.style={
        /pgfplots/error bars/draw error bar/.code 2 args={%
            \pgfkeysgetvalue{/pgfplots/error bars/error mark}%
            {\pgfplotserrorbarsmark}%
            \pgfkeysgetvalue{/pgfplots/error bars/error mark options}%
            {\pgfplotserrorbarsmarkopts}%
            \path ##1 -- ##2;
        },
        /pgfplots/table/.cd,
        y index=4,
        y error expr={\thisrowno{2}-\thisrowno{4}},
        /pgfplots/box plot
    },
    box plot bottom whisker/.style={
        /pgfplots/error bars/draw error bar/.code 2 args={%
            \pgfkeysgetvalue{/pgfplots/error bars/error mark}%
            {\pgfplotserrorbarsmark}%
            \pgfkeysgetvalue{/pgfplots/error bars/error mark options}%
            {\pgfplotserrorbarsmarkopts}%
            \path ##1 -- ##2;
        },
        /pgfplots/table/.cd,
        y index=5,
        y error expr={\thisrowno{3}-\thisrowno{5}},
        /pgfplots/box plot
    },
    box plot median/.style={
        /pgfplots/box plot
    }
}
\begin{document} 
\title{Early Identification of Services in HTTPS Traffic}


\author{\IEEEauthorblockN{
Wazen M. Shbair \IEEEauthorrefmark{1},
Thibault Cholez \IEEEauthorrefmark{2},
J\'er\^ome Fran\c cois \IEEEauthorrefmark{3},
Isabelle Chrisment \IEEEauthorrefmark{2}
}

\IEEEauthorblockA{
\IEEEauthorrefmark{1} University of Luxembourg, SnT, 29, Avenue J.F Kennedy, L-1855 Luxembourg\\
Email:\{wazen.shbair\}@uni.lu\\
\IEEEauthorrefmark{2} University of Lorraine, LORIA, UMR 7503, Vandoeuvre-les-Nancy, F-54506, France\\ 
\IEEEauthorrefmark{3} INRIA Nancy Grand Est, 615 rue du Jardin Botanique, 54600 Villers-les-Nancy, France\\
Email: \{thibault.cholez, jerome.francois, isabelle.chrisment\}@loria.fr\\
 }
  }


\IEEEtitleabstractindextext{%
\begin{abstract}
Traffic monitoring is essential for network management tasks that ensure security and QoS. However, the continuous increase of HTTPS traffic undermines the effectiveness of current service-level monitoring that can only rely on unreliable parameters from the TLS handshake (X.509 certificate, SNI) or must decrypt the traffic. We propose a new machine learning-based method to identify HTTPS services without decryption. By extracting statistical features on TLS handshake packets and on a small number of application data packets, we can identify HTTPS services very early in the session. Extensive experiments performed over a significant and open dataset show that our method offers a good accuracy and a prototype implementation confirms that the early identification of HTTPS services is satisfied.

\end{abstract}

}

\maketitle

 \IEEEdisplaynontitleabstractindextext 

%
\IEEEpeerreviewmaketitle

\section{Introduction}
\IEEEPARstart{E}{arly} identification of Internet traffic is vital for network operators. They need to know what is passing through their network as early as possible thanks to their monitoring infrastructure, so that they can take the proper actions to manage the network, as for example for security, QoS or traffic engineering purposes. Indeed, observing relevant events to these tasks has to be done as soon as possible to avoid any security breach, malfunction or downtime. Therefore, network monitoring and analysis tools must be continuously adapted to pursue the quick evolution of the network traffic and to cope with modern challenges.

In particular, today's network monitoring solutions are losing their power due to the global trend toward encryption of network communications. 
Encryption undermines the effectiveness of standard monitoring approaches and makes it difficult to identify and monitor the services behind encrypted traffic, which is essential to properly manage the network. On the web,  Transport Layer Security (TLS) is the most widely used encryption protocol. 
The HTTPS protocol allows web applications to secure HTTP over TLS, which makes it difficult for a third party to infer information about users' interaction with a website using packets sniffer or Man-in-the-Middle attacks \cite{McCarthy2011}. 

The trend of using HTTPS protocol has created an "\textit{Encryption Rush}" on Internet industry in two dimensions: the size of HTTPS traffic, and the number of HTTPS websites. According to Cisco 2018 annual security report \cite{Cisco2018} 50\% of global web traffic was encrypted as of October 2017. That is a 12-point increase in volume from November 2016.
Also the augmentation of HTTPS traffic is recorded by Mozilla telemetry, where it shows that the HTTPS traffic has reached a tipping point and passed the halfway mark of the total size of web traffic \cite{mozilla2016}.  


But HTTPS is a double edged sword. On one side, the increasing share of the HTTPS traffic has been a mostly positive step toward more security and privacy on the web. On the other side, HTTPS does not prevent malicious websites despite the check of SSL certificates \cite{Kaspersky2014}. The authors in \cite{chen2014security} found that 13\% of Chinese websites 
are using self-signed SSL certificates instead of one issued and verified by a Certificate Authority (CA). Even worse, based on \cite{pukkawanna2014classification}, there is a significant number of phishing websites that use valid SSL certificates issued by trusted CAs to convince clients to trust them. Therefore, there is a high demand for solutions able to monitor HTTPS traffic. 
Three practical methods are used to early identify HTTPS services, namely: Server Name Indication (SNI), SSL certificate and HTTPS proxy. In our previous work \cite{shbair2015efficiently} the reliability issues of the first two approaches were explained. Concerning the HTTPS proxy, it decrypts HTTPS traffic between a client and server at the cost of huge trust and privacy issues and can only be used in very sensitive contexts. 

Hereby, we present a new approach to identify HTTPS services very early without any decryption and without relying on simple protocol headers' values but instead on the traffic pattern itself learned from the first few packets. Early identification means are at the core of middle-boxes like Network Intrusion Detection Systems (NIDS) to identify and manage services, for example to prevent access to unsafe websites. Also, it can be used to apply QoS by triggering automated re-allocation of network resources for high priority services such as webmail \cite{nguyen2008survey}.

The contribution of this paper is fourfold: (1) a novel processing of the TLS handshake and few application data packets to identify services in HTTPS traffic very early in the session. (2) a comprehensive HTTPS identification framework, that is able to monitor HTTPS services without decryption. (3) the evaluation of the efficiency of our solution over an open HTTPS dataset. (4) a prototype software that demonstrates the early identification and practical applicability of the proposed approach. While our previous work \cite{shbair2016multi} was focused on the definition of the statistical features and of the multi-level identification approach, this works extends and refines our initial contribution to make it work with only a few packets instead of the full flow, what widens its applicability.

The rest of the text is organized as follows:
\Cref{sec:related-work} explores related work on HTTPS identification. Our methodology is presented in \Cref{sec:realtimeIde}. \Cref{sec:evaluationex} evaluates the proposed approach, features, algorithm and the used dataset. \Cref{sec:realprotoype} presents our proof-of-concept prototype, it also discusses the prototype performance for identifying HTTPS service. Finally, \Cref{sec:conclusion} concludes the paper.


\section{Related Work}
\label{sec:related-work}
Identifying network traffic on the fly with machine learning has two main challenges before even discussing the statistical features and the method to use: reassembling flows and gathering a sufficient number of packets needed to classify a flow. In this section, we explore the literature work that discusses these issues.

\textit{Flows reassembly}: a TCP flow is defined as a set of packets that have the same 4-tuples (source IP address, source port, destination IP address and destination port) \cite{ong2009}. The HTTPS flow contains three types of packets: Handshake packets, Application data packets and pure TCP control packets. The idea of flow demultiplexing is to build a table (e.g., Hash table), where each record contains the packets from a single TCP flow and is accessed by the key of the flow (i.e., the 4-tuple). For each new incoming packet, the 4-tuple key is computed and used for searching in a table to find the corresponding flow of the packet. 

Xiong et al. \cite{ong2009} propose a TCP flow reassembly approach for real-time network traffic processing in high-speed networks. They used the \textit{Recently Accessed First} principle to reduce the search cost of the related flow of incoming packets, by bringing the most recently accessed flow record on top of the table, so the next following packets will be mapped quickly. In \cite{PEREIRA2014} the authors also apply the same method for flow reassembly: they run packet capture, flow reassembly, and classification with machine learning modules in a pipeline way. Their method achieves a reassembly throughput of 24,997.25 flows/second, while the average delivery delay is 0.49 seconds. Groleat et al. \cite{groleat2014high} also share the idea of building a real-time classifier based on SVM algorithm. They tried to detect the categories of network applications, while we aim to be more precise to identify flows at service-level. In their work, they proposed a high-speed flow reassembly algorithm able to handle one million concurrent flows by using massive parallelism and low-level network interface access of Field-Programmable Gate Array (FPGA) boards. Their results show up to 20 GB/s for flow reassembly. In our work, we benefit from the approach in \cite{ong2009} to optimize the HTTPS flow reassembly module. The flow reassembly using hardware facility like FPGA is promising, but we keep it for our future work.


\textit{Sufficient Number of Packets}: the sufficient number of packets is a critical parameter, since packets are required to obtain statistics about the flows that are used as input for identification methods. Many research works have investigated how many packets are actually needed to obtain sufficient statistics to keep high accuracy levels. Bernaille et al. \cite{bernaille2007early} report using the first 4 packets to early identify applications over TLS traffic like FTP, SMPT or POP3. Kumano et al. \cite{kumano2014towards} investigate the real-time identification of applications type (e.g., P2P, Streaming, etc.) in encrypted traffic. Their results show that they need 70 packets for their real-time identification. Maolini et al. \cite{MAIOLINI2009} handle the problem of identifying SSH traffic and the underlying protocols (SCP, SFTP and HTTP). Their results show that the first 3 to 7 packets with a $K$-means clustering algorithm are sufficient to achieve a real-time identification. In \cite{Yanai2010} the authors use the first 100 packets for on the fly classification between applications like SMTP, POP3, eDonkey and BitTorrent. They have combined $K$-means and $K$-nearest neighbour clustering algorithms to build a lightweight classifier. 

Based on the aforementioned studies, we can see that values range from 3 to 100 packets. In our case, we can not clearly deduce from this state-of-the-art the sufficient number of packets we need, since we have a more precise target (i.e., service-level identification), which is more challenging than the application-level identification of the aforementioned studies that consider HTTPS traffic as one class. Therefore, we have to evaluate carefully the sufficient number of packets that makes it possible to early recognize HTTPS services.

\textit{Early Identification of Network Flows:} 
intuitively, offline identification methods should give better results since they depend on stable and complete flows' statistics. In offline approaches, the identification process starts after network flows have been terminated and only then statistical features are computed regardless of time and computation complexity. However, the problem is harder in the context of early identification, where the identification runs in parallel with flows and the goal is to name the HTTPS service behind a live flow as soon as possible, by using as little information as possible without sacrificing accuracy. 


Many efforts have been made to use machine learning techniques and to make them efficient for early identification. The authors in \cite{Jun2008, Yanai2010, Chengjie2011, PEREIRA2014, kumano2014towards} apply different machine learning approaches (C4.5, REPTree, K-means, SVM, Naive Bayes) for identifying the application type (e.g., P2P, SSH, BitTorrent) of encrypted traffic in real-time manner with acceptable level of accuracy and overhead. We are going much further into the identification granularity by early identifying specific HTTPS services.

\section{Early Identification of HTTPS Services}
\label{sec:realtimeIde}

The proposed framework contains two phases: the first phase includes offline learning and model building and the second phase concerns real-time flow reassembly and features extraction. For the early identification, we use TLS handshake packets and only the first application data packets. The reasons behind this choice are: First, to minimize the overhead of the flow reassembly step, as it should reassemble only a few packets. In the rest of this paper, we mean by flow reassembly the gathering of the TLS handshake packets and a small number of application data packets. Second, to decrease the effect of out-of-order packets, which is a challenge in early identification systems \cite{Bernaille2006}. We benefit from the fact that the TLS handshake messages must arrive in-order to establish the TLS connection. Third, to 
 limit the sensibility of the signature to services' evolution. The continuous evolution of web applications creates an overhead to machine learning based identification methods that must re-evaluate their statistical features periodically to cope with changes. Thus, using TLS handshake messages and only few application data packets should reduce the effect of traffic characteristics changes in time, since handshake messages are only used to establish TLS connections. To our knowledge, it is the first time that TLS handshake and few application data packets are used to identify HTTPS services.





Formally, the identification of HTTPS services can be described as follows. Let us assume $X=\{ x_{1},x_{2},...,x_{n}\}$ a set of HTTPS flows. A flow instance $x_{i}$ contains $h$ TLS handshake packets, and $d$ application data packets. A flow $x_{i}$ is characterized by a vector of $r$ features (e.g., packets mean size). Also let $Y=\{y_{1},y_{2},..., y_{s}\}$ be the set of known HTTPS services, where $s$ is the total number of known HTTPS services. The $y_{i}$ is the HTTPS service behind an HTTPS flow $x_{i}$. Thus, our target is to learn the mapping of a $r-$ dimensional variable $X$ to $Y$. 

\subsection{The statistical features} 
\label{sec:machine}
Identifying HTTPS services on the fly requires to limit the choice of features to those that can be computed quickly on partial flows. In addition, a threshold must be determined that is the sufficient number of packets that must have been captured from a flow to start the extraction of features and perform the identification \cite{Yanai2010}. The proper threshold value will be discussed and evaluated later. 

According to our public HTTPS dataset (presented in \Cref{sec:evaluationex}), the number of handshake packets is not fixed among TLS sessions, since the TLS handshake can take three or four packets, or even more depending on whether a client and a server start a new fresh handshake or resume a previous one. Therefore, we use the start of exchanging application data packets as the indicator of the handshake phase completion. Then, we further consider the first application data packets.

Hence, a set of statistical features that contains three group of features has been defined; the first group is common to all packets (i.e., TLS handshake and application data packets), the second one is related to the TLS handshake headers , while the third group is related to application data packets. Table \ref{features} summarized the 36 features we have selected for identifying HTTPS services in real-time. Some of these features were evaluated in our previous work \cite{shbair2016multi}. 

\begin{table}[]
\setlength{\belowcaptionskip}{-10pt}
\centering
\footnotesize
\caption{Our 36 statistical features over TLS handshake and application data packets}
\label{features}
\begin{tabular}{|c|}
\hline
\textbf{\begin{tabular}[c]{@{}c@{}}Common features (9 features per direction)\end{tabular}} \\ \hline
\begin{tabular}[c]{@{}c@{}}Packet size (Average, 25th, 50th, 75th Percentile, Variance, Maximum)\\ Inter arrival time (25th,50th,75th Percentile)\end{tabular} \\ \hline \hline
\textbf{\begin{tabular}[c]{@{}c@{}}TLS handshake header features (6 features)\end{tabular}} \\ \hline
\begin{tabular}[c]{@{}c@{}}\texttt{ClientHello} packet\\ (length of Session ID, number of cipher suites, length of extensions list)\end{tabular} \\ \hline
\begin{tabular}[c]{@{}c@{}}\texttt{ServerHello} packet \\ (length of Session ID, used cipher suite, length of extension list)\end{tabular} \\ \hline \hline
\textbf{\begin{tabular}[c]{@{}c@{}}Application data packets features (6 features per direction)\end{tabular}} \\ \hline
Packet size (Average, 25th, 50th, 75th Percentile, Variance, Maximum) \\ \hline
\end{tabular}
\end{table}

\subsection{Selected machine learning algorithm}
\label{sec:ml}
Once the features are computed, they are used as input for a machine learning algorithm. Many machine learning algorithms have been already used for encrypted traffic classification, such as C4.5, Naïve Bayes, SVM, etc. All machine learning algorithms identify traffic by using features as input of their classification method. There is no best classification model for all applications but some may perform better than others in our case. For instance, according to \cite{xue_traffic_2013}, if we want to differentiate traffic generated by multiple protocols, the C4.5 algorithm would be better than SVM algorithm, but if we target a small number of network applications, the SVM would be more suitable. 

In the field of encrypted traffic classification some machine learning algorithms have been used extensively. The authors in\cite{alshammari2009machine} used Naïve Bayesian, RIPPER and C4.5 for detecting SSH and Skype traffic. McCarthy et al. \cite{mccarthy_investigation_2011} used C4.5, RIPPER, and Naïve Bayes techniques to detect TLS traffic, while Li et al. apply the RandomForest algorithm \cite{li2007identifying}. In \cite{singh2013near} the authors find that C4.5 and Naïve Bayes provide a good accuracy in the context of a near real-time identification of network traffic. In this work, we only consider the C4.5 algorithm. The main reason is based on the performance comparison between different machine learning algorithms conducted in \cite{Williams2006} for a similar use case and where the results show that the C4.5 algorithm is very fast and efficient, what becomes particularly important in the context of early identification. It also has a low computational overhead, simple to implement and it requires less learning time and classification time \cite{Jun2008}.

\subsection{Architecture}
\begin{figure*}[]
\centering
\includegraphics[scale=0.6]{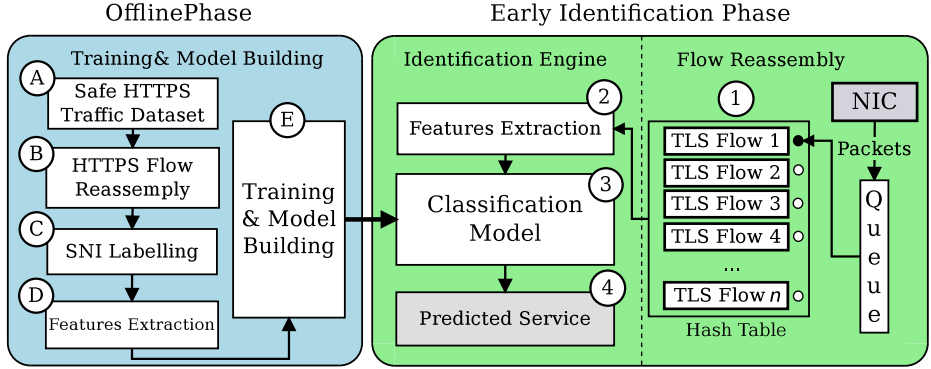}
\caption{Framework architecture for early identification of HTTPS services}
\label{realtimeframework}
\end{figure*}

Figure \ref{realtimeframework} illustrates the phases of the proposed approach. First, we run the offline part to build the classification model, which will be used in the early identification phase for identifying HTTPS services. The offline phase has been detailed extensively in our previous work \cite{shbair2016multi}.
The early identification phase consists in the flow reassembly process and the identification engine. The core of the identification engine is prepared in the offline phase. The flow reassembly module demultiplexes the TLS handshake and application data packets into the corresponding flow. 
Figure \ref{realtimeframework} illustrates the interactions of the proposed approach in the early identification phase as follows:
(1) The reassembly module receives TLS packets from the Network Interface Card (NIC) and demultiplexes incoming packets, (2) When the number of packets in a flow reaches a given threshold, the feature extraction over the flow packets is triggered, (3) Computed features are given to the pre-built classification model for identification and (4) Finally, the predicted HTTPS service name is given. 

\label{sec:arch}


\section{Evaluation and Experimental Results}
\label{sec:evaluationex}
This section evaluates the key parameters that may affect the accuracy of the proposed solution: the HTTPS dataset and the threshold value that defines the sufficient number of packets needed to early identify HTTPS services with a good accuracy. For the evaluation, we use the features set described in \Cref{sec:machine} with the C4.5 machine learning algorithm. Except if explicitly mentioned, we use the 10-fold cross validation (i.e, 90\% of data for training and 10\% for testing). Our main evaluation metric is the {Accuracy} defined as: \textit{the percentage of correctly classified instances over the total number of instances }\cite{Williams2006}.

\subsection{Public HTTPS dataset}
\label{sec:datasetcollection} 
We built our own dataset due to the the lack of available HTTPS traces dataset including packet's payload. Thanks to an automated script, we visited the top 1000 HTTPS websites referenced in the Internet-Wide Scan Data Repository\footnote{\url{https://scans.io/series/443-https-tls-alexa}} on a daily basis during two weeks, using both Google Chrome and Firefox web browsers. Our script  loads the main page of each website, which is sufficient to get the beginning of the flows and does not create privacy issues when releasing the dataset, since no real users generated the traffic.

The dataset we made contains 487312 flows related to 7977 different HTTPS services for a total size of 55 GB. This higher number of services is due to many HTTPS flows being from third-party content providers that are used to render the requested web pages: load the website content, display advertisement, make statistics, etc. The collection process took place over two weeks. Therefore, we consider only the HTTPS services that appeared at least 14 times (one per day) in our traces and are included in the original target list. It means, for each HTTPS service, a minimum number of 14 labelled flows per service is needed to train the classification model. The dataset has been pre-processed by eliminating uncompleted flows with no application data packets. We indeed need to be sure that we use valid flows to train our model.

The dataset is divided in two parts according to the web browser used to access the websites. Some flows were issued by Google Chrome (70771 labelled flows related to 1384 HTTPS services), and the others by Firefox (56116 labelled flows related to 890 HTTPS services). Therefore, we evaluate our approach over HTTPS services accessed by Google Chrome and Firefox to be sure that the classification results are not sensitive to the client but just to the accessed service. 

For the sake of reproducibility, our HTTPS dataset is publicly available \footnote{\url{http://betternet.lhs.loria.fr/datasets/https/}}
 with full encrypted payloads and the TLS layer information. We hope to contribute in solving the absence of reference datasets for HTTPS \cite{Velan2015}.

\subsection*{Overview of the HTTPS dataset}
\begin{figure}[hbtp]
\setlength{\belowcaptionskip}{-10pt}
\centering
\begin{tikzpicture}
\begin{axis}[ 
	height=4.5cm, ylabel near ticks,
	width=9cm,bar width=10pt,
	tick label style={font=\scriptsize},
	label style={font=\scriptsize},
	xlabel= Number of TLS handshake packets,
	ylabel= Proportion of HTTPS flows,style={font=\scriptsize},
	ybar, ymax=0.8,
    legend style={font=\scriptsize},  
	xtick={0,3,4,5,6},
	symbolic x coords={0,3,4,5,6} 
]

\addplot [ybar,fill=yellow, postaction={pattern=north west lines}] coordinates{
(3,0.21) (4,0.75) (5,0.03)};
\addplot [ybar,fill=green, postaction={pattern=north east lines}] coordinates{
(3,0.32) (4,0.66) (5,0.1)};

\legend{Google Chrome, Firefox}
\end{axis}
\end{tikzpicture}
\caption{Distribution of the number of TLS handshake packets per flow in the dataset}
\label{handshekdist}
\end{figure}
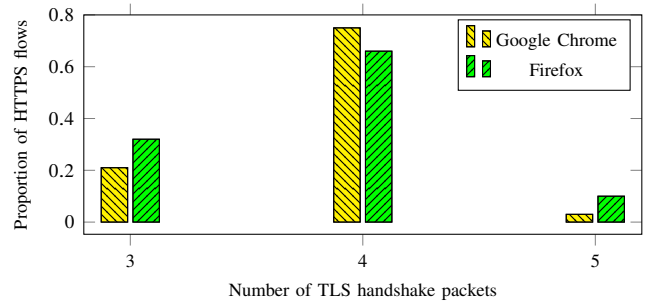

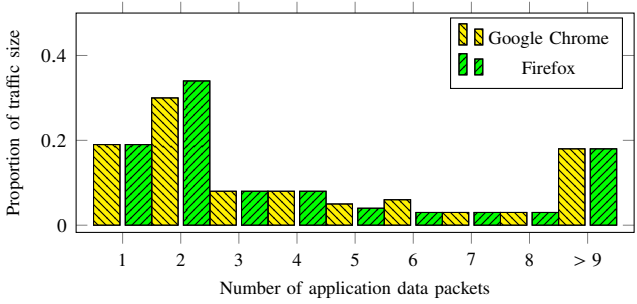
\begin{figure}[hbtp]
\setlength{\belowcaptionskip}{-5pt}
\centering
\begin{tikzpicture}
	\begin{axis}[
		height=4.5cm, bar width=10pt,ylabel near ticks,
		width=9cm, 
		xlabel= Number of application data packets,
		ylabel= Proportion of traffic size, 
		tick label style={font=\scriptsize},
		label style={font=\scriptsize},
		ybar ,ymax=0.5,legend style={font=\scriptsize}, 
		xtick={1,2,3,4,5,6,7,8,$>9$},
		symbolic x coords={1,2,3,4,5,6,7,8,$>9$},
		 	]
\addplot[ybar,fill=yellow, postaction={pattern=north west lines}] coordinates{
(1,0.19)(2,0.3)
(3,0.08)(4,0.08)
(5,0.05)(6,0.06)
(7,0.03)(8,0.03)
($>9$,0.18)
};

\addplot[ybar,fill=green, postaction={pattern=north east lines}] coordinates{
(1,0.19) (2,0.34) (3,0.08) (4,0.08) (5,0.04) (6,0.03)
(7,0.03) (8,0.03) ($>9$,0.18)};

\legend{Google Chrome, Firefox}
\end{axis}
\end{tikzpicture}
\caption{Distribution of the number of application data packets per flow in the dataset}
\label{apppacketshisto}		
\end{figure}

As explained before, our approach relies on features calculated over the TLS handshake packets and over application data packets. This section explores the number of TLS handshake packets and of application data packets present in the different flows that constitute our HTTPS dataset. 
Figure \ref{handshekdist} depicts that the number of TLS handshake packets is between 3 and 5 packets. We can notice close yet different distributions between Google Chrome and Firefox but the majority of flows have 4 TLS handshake packets, 75\% for Google Chrome and 66\% for Firefox. Regarding application data packets, Figure \ref{apppacketshisto} shows the distribution of flows according to the total number of application data packets they have. There are many short HTTPS flows with only one or two application data packets (49\% for Google Chrome, 53\% for Firefox), while a relatively small number of flows are long with more than 9 application data packets (18\% for Google Chrome, 18\% for Firefox), even if they represent more in terms of the proportion of data exchanged.

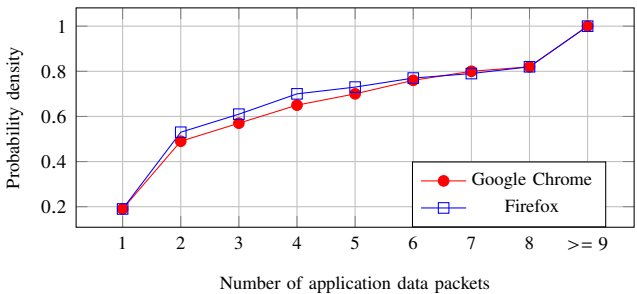
\begin{figure}[hbtp]
\setlength{\belowcaptionskip}{-10pt}
\centering
\begin{tikzpicture}
\begin{axis}[ 
	height=4.5cm, ylabel near ticks,
	width=9cm,
	tick label style={font=\scriptsize},
	label style={font=\scriptsize},
	xlabel= Number of application data packets,
	ylabel= Probability density,style={font=\scriptsize},
	grid=major,xtick=data,
	symbolic x coords={1,2,3,4,5,6,7,8,$>=9$},
	legend style={at={(0.95,0.3)}}
]
\addplot[color=red,mark=*] coordinates{
(1,0.19) (2,0.49)
(3,0.57) (4,0.65)
(5,0.70) (6,0.76)
(7,0.80) (8,0.82)
($>=9$,1)
};
\addplot[color=blue,mark=square]  coordinates{
(1,0.19) (2,0.53)
(3,0.61) (4,0.70)
(5,0.73) (6,0.77)
(7,0.79) (8,0.82)
($>=9$,1)
};
\legend{Google Chrome, Firefox}
\end{axis}
\end{tikzpicture}
\caption{CDF for the number of application data packets per flow}
\label{cdf}        
\end{figure}

Figure \ref{cdf} depicts the Cumulative Distribution Function (CDF) of the number of application data packets per flow. A significant proportion of flows (49\%) consists of one or two application data packets, while flows with 9 or more application data packets account for about 18\%. From another point of view, Table \ref{flowsize}, shows the contribution of the different flows regarding their number of application data packets to the total size of the dataset. The flows with one application data packet account for 15.3\% of the total traffic size with Google Chrome, and 14.13\% with Firefox. It can be noticed that short flows (i.e., with less than 9 application data packets) account for almost two thirds of the total size ($\simeq 64\%$) while long flows account for a bit more than the last third ($\simeq 35\%$)). 
This can be explained by the specific nature of web traffic that vehiculates a lot of small objects like CSS files, HTML titles, JavaScript, etc. So these short TCP flows are almost related to simple HTTP GET and POST requests used to get the web pages' content \cite{qian2012flow}. These statistics reflects the way the dataset was generated (i.e. by loading the front page of each website) and may not be representative of HTTPS traffic at a larger scale which may hold other kinds of traffic, for example large sessions of video streaming over HTTPS. Anyway, our goal being to identify services by using very few packets, the dataset is perfectly adapted to our objective.

\begin{table}[]
\setlength{\belowcaptionskip}{-5pt}
\caption{The participation in the dataset size per flow size}
\centering
\begin{tabular}{|c|c|c|}
\hline
\begin{tabular}[c]{@{}c@{}} Application \\ Data Packets\end{tabular} & Google Chrome & FireFox \\ \hline
1 & 15\% & 14\% \\ \hline
2 & 19\% & 16\% \\ \hline
3 & 8\% & 7\% \\ \hline
4 & 7\% & 7\% \\ \hline
5 & 5\% & 5\% \\ \hline
6 & 5\% & 6\% \\ \hline
7 & 4\% & 4\% \\ \hline
8 & 4\% & 6\% \\ \hline
\textgreater{}9 & 34\% & 35\% \\ \hline
\end{tabular}
\label{flowsize}

\end{table}


    
\vspace{-0.5cm}
\subsection{Evaluation of the number of application data packets characterizing a flow}
\label{sec:apppacketsnum}

As shown in the previous section, the majority of flows have 3 to 4 TLS handshake packets, however there is a large variety in the number of application data packets per flow. Hence, our objective is to use the minimum number of application data packets that enables an accurate early identification of HTTPS services. So we evaluate in this section the correlation between the accuracy and the number of application data packets considered.
 
First, long HTTPS flows ($>$ 15 data packets) are handled. The target is to assess the accuracy to identify HTTPS services behind such long flows using different $d$ numbers of application data packets, where $ d \in \{0,1,2,3,4,5,6,9,12,15\}$. Hence, 10 classification models were built. Each model was created using all the TLS handshake packets and $d$ application data packets. For example, the first model consists in considering only the TLS handshake packets ($d=0$) to build the model and identify HTTPS services, while the second model includes TLS handshake packets plus one data packet ($d=1$).

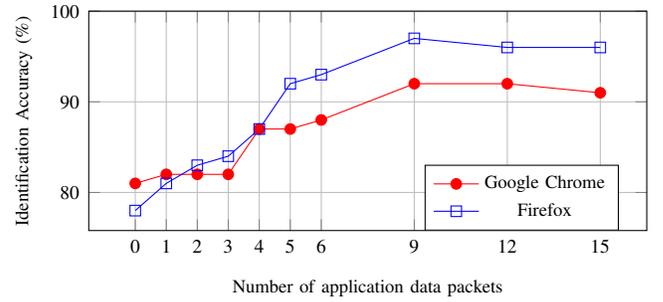
\begin{figure}
\setlength{\belowcaptionskip}{-5pt}
\centering
\begin{tikzpicture}
\begin{axis}[ 
	height=4.5cm, ylabel near ticks,
	width=9cm,
	tick label style={font=\scriptsize},
	label style={font=\scriptsize},
	xlabel= Number of application data packets,
	ylabel= Identification Accuracy (\%),style={font=\scriptsize},
	grid=major,xtick=data,
	ymax=100, 
	legend style={at={(0.95,0.3)}}
]
\addplot[color=red,mark=*] coordinates{
(0,81) (1,82) (2,82)
(3,82) (4,87) (5,87)
(6,88) (9,92) (12,92)(15,91)
};
\addplot[color=blue,mark=square]  coordinates{
(0,78) (1,81) (2,83)
(3,84) (4,87) (5,92)
(6,93) (9,97) (12,96)
(15,96)
};
\legend{Google Chrome, Firefox}
\end{axis}
\end{tikzpicture}
\caption{Impact of the number of application data packets considered on the identification accuracy, when identifying \underline{long flows} ($>$ 15 data packets per flow)}
\label{services15}        
\end{figure}

In the dataset, we find 171 different HTTPS services accessed with Google Chrome and 124 HTTPS services accessed with Firefox having more than 15 application data packets. Figure \ref{services15} shows the relation between the number of application data packets considered by the model and the overall identification accuracy. As expected, increasing the number of application data packets improves the accuracy. However, after 9 application data packets the identification accuracy is almost stable, for Google Chrome we get $91.66\%\pm0.57$, and for Firefox $96.33\%\pm0.47$. Thus, it is possible to identify HTTPS services behind long flows quite early by using only the first 9 application data packets and without decreasing the accuracy.

\begin{figure}
\setlength{\belowcaptionskip}{-5pt}
\centering
\begin{tikzpicture}
\begin{axis}[ 	
	height=4.5cm, ylabel near ticks,
	width=9cm,
	tick label style={font=\scriptsize},
	label style={font=\scriptsize},
	xlabel= Number of application data packets,
	ylabel= Identification Accuracy (\%),style={font=\scriptsize},
	grid=major,xtick=data,
	ymax=100, 
	legend style={at={(0.95,0.3)}}
]
\addplot[color=red,mark=*] coordinates{
(0,79.9) (1,90.25) (2,90.72)(3,90.77) (4,91.27) (5,91.31)(6,91.40) (7,91.55) (8,91.50) (9,91.67)};
\addplot[color=blue,mark=square]  coordinates{
(0,77.4)(1,90.2)(2,91.2)(3,92.3)(4,91.5)(5,92)(6,92.04)(7,92)(8,92)(9,92.3)};
\legend{Google Chrome, Firefox}
\end{axis}
\end{tikzpicture}
\caption{Impact of the number of application data packets considered on the identification accuracy when identifying \underline{all flows}}
\label{allservicesincluded}        
\end{figure}
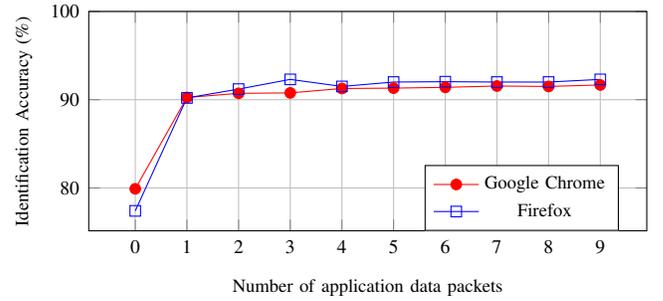

However, as shown in Figure \ref{apppacketshisto}, most flows of the dataset are short (82\% of flows have less than 9 application data packets).
Thus, a classification model should be able to handle various number of application data packets to be able to identify all flows. Figure \ref{allservicesincluded} depicts the performance of C4.5 models that have been built by progressively increasing the maximum number of application data packets considered when building/training the model and applied to identify all HTTPS services considered in the dataset (1384 of Google Chrome and 890 of Firefox) 
irrespectively of the flow size.
For instance, if we use up to three application data packets, HTTPS flows with one, two or three data packets are fully given for training and testing, while longer flows are limited to the three first application data packets. It can be noticed that with no application data packets (only handshake packets), the classification of Google Chrome and Firefox traffic achieves 79.9\% and 77.4\% identification accuracy respectively. However, the accuracy can be increased to 90.25\% and 90.2\% if a single application data packet is added. If we include up to 4 application data packets, the accuracy rises respectively to 91.27\% and 91.5\%. 

\vspace{-0.3cm}
\subsection*{Decoupling the number of application packets used in training phase from testing phase}
\label{milestonevalue}

A deeper analysis is conducted to assess the performance of the proposed method by measuring the identification accuracy with less or more application data packets when trying to identify a service than the number used for training. 

Varying the number of application data packets is valid from the ML-algorithm point of view because we use flow-level features and not packet-level features. So, the space used to perform the classification is the same and the number of features used does not vary with the number of application packets. For example, computing the maximum packet-size of a flow over 3 or 5 packets only counts as a single feature. But the measured value of the feature may change regarding the number of input packets used for its computation.

Let's assume, we train a classification model to identify a given HTTPS service $X$ using 4 handshake packets plus up to $d$ application data packets. This means that features related to application data packets (see Table \ref{features}), will be computed over $d$ or less application data packets. However, we deal with application data packets, which  number is not fixed and varies according to the amount of data to be exchanged. So if the model is given features from a flow $X$ computed over 4 handshake packets but with $r$ application data packets, where $r \neq d$, we need to assess how much the accuracy of the model will be affected. The motivation is to explore what is the best number of packets $d$ to consider when building the model and how another identification threshold (fewer or higher) can affect (or not) the accuracy.

In this part, our approach is still validated using flows accessed by Google Chrome and Firefox. Table \ref{onemodelforall} shows the accuracy using a C4.5 model with up to $d$ application packets, where $ d \in \{1,2,3,4,5,6,9\}$ to identify HTTPS services accessed by Google Chrome, and Table \ref{firefoxforall} for the ones accessed by Firefox. 
Typically, the interesting result is to see how a short training threshold affects the identification accuracy of long flows and how a high threshold affects the identification of short flows, in order to find a good trade-off. Please note that 10-fold cross validation was \underline{not} used for this experiment to limit the noise. Consequently, when training and testing thresholds are aligned, the accuracy is always of 100\% since we train and test on the same flows (i.e. all flows are used for training, no flows are kept for testing). But this result is not interesting and has been removed from the tables. It is also redundant with the experiment leading to Figure \ref{allservicesincluded} that used 10-fold cross validation and aligned training thresholds. 
 
In the first row in Table \ref{onemodelforall}, we build a C4.5 classification model by calculating the statistical features over the TLS handshake packets plus a single application data packet. Then this model is tested to identify flows by looking at different numbers of application data packets. The idea is to assess the identification accuracy by using only one application data packet for training. For instance the second column (in the first row) shows an accuracy of 96\% for identifying flows by looking at their first two application data packets. The last column (still in the first row) gives the identification accuracy 90\% for identifying flows with up to 9 application data packets, with the same training configuration. This table shows that with a low training threshold the accuracy decreases when the testing threshold increases, but with a high training threshold, accuracy decreases even faster when the testing threshold lowers. 


According to these results (for both Google Chrome and Firefox), there is a trade-off between the number of allowed application data packets and the level of identification accuracy. Thus, we conclude that using a training model that has been built using up to 4 or 5 application data packets offers more balanced results to identify both short and long HTTPS flows. As shown in Table \ref{onemodelforall} and Table \ref{firefoxforall} using up to 5 application data packets achieves (75\%, 73\%) identification accuracy with HTTPS flow with a single application data packet and (94\%, 96\%) to identify ones with 9 application data packets for Google Chrome and Firefox respectively.


\begin{table*}[]
\setlength{\belowcaptionskip}{-5pt}
\footnotesize
\centering
\caption{Identifying HTTPS services accessed by Google Chrome}
\label{onemodelforall}
\begin{tabular}{|l|l|c|c|c|c|c|c|c|c|c|}
\hline
\multicolumn{2}{|l|}{\multirow{2}{*}{}} & \multicolumn{9}{c|}{Testing (up to $d$)} \\ \cline{3-11} 
\multicolumn{2}{|l|}{} & 1 & 2 & 3 & 4 & 5 & 6 & 7 & 8 & 9 \\ \hline
\multirow{9}{*}{\begin{tabular}[c]{@{}l@{}}\rot{Training (up to $d$)}\end{tabular}} & 1     
 & - & 96\% & 94\% & 91\% & 90\% & 90\% & 90\% & 90\% & 90\% \\ \cline{2-11} 
 & 2 & 83\% & - & 95\% & 90\% & 89\% & 90\% & 89\% & 88\% & 87\% \\ \cline{2-11} 
 & 3 & 80\% & 95\% & - & 95\% & 94\% & 93\% & 92\% & 92\% & 91\% \\ \cline{2-11} 
 & 4 & 72\% & 92\% & 96\% & -& 98\% & 97\% & 96\% & 94\% & 93\% \\ \cline{2-11} 
 & 5 & 75\% & 89\% & 94\% & 97\% & - & 98\% & 97\% & 95\% & 94\% \\ \cline{2-11} 
 & 6 & 73\% & 88\% & 92\% & 96\% & 98\% & - & 99\% & 97\% & 96\% \\ \cline{2-11} 
 & 7 & 74\% & 88\% & 92\% & 95\% & 97\% & 99\% & - & 98\% & 97\% \\ \cline{2-11} 
 & 8 & 73\% & 88\% & 92\% & 95\% & 96\% & 98\% & 99\% & - & 99\% \\ \cline{2-11} 
 & 9 & 71\% & 86\% & 93\% & 95\% & 96\% & 97\% & 98\% & 99\% & - \\ \hline
\end{tabular}
\end{table*} 

\begin{table*}[]
\setlength{\belowcaptionskip}{-5pt}
\footnotesize
\centering
\caption{Identifying HTTPS services accessed by Firefox}
\label{firefoxforall}
\begin{tabular}{|l|l|c|c|c|c|c|c|c|c|c|}
\hline
\multicolumn{2}{|l|}{\multirow{2}{*}{}} & \multicolumn{9}{c|}{Testing (up to $d$)} \\ \cline{3-11} 
\multicolumn{2}{|l|}{} & 1 & 2 & 3 & 4 & 5 & 6 & 7 & 8 & 9 \\ \hline
\multirow{9}{*}{\begin{tabular}[c]{@{}l@{}}\rot{Training (up to $d$)}\end{tabular}} & 1 & - & 84\% & 79\% & 77\% & 76\% & 76\% & 76\% & 76\% & 76\% \\ \cline{2-11} 
 & 2 & 83\% & - & 91\% & 84\% & 81\% & 80\% & 79\% & 78\% & 78\% \\ \cline{2-11} 
 & 3 & 76\% & 93\% & - & 96\% & 95\% & 93\% & 92\% & 91\% & 90\% \\ \cline{2-11} 
 & 4 & 73\% & 88\% & 96\% & - & 98\% & 96\% & 95\% & 94\% & 93\% \\ \cline{2-11} 
 & 5 & 73\% & 88\% & 94\% & 97\% & - & 98\% & 97\% & 96\% & 96\% \\ \cline{2-11} 
 & 6 & 70\% & 85\% & 91\% & 96\% & 98\% & - & 99\% & 98\% & 98\% \\ \cline{2-11} 
 & 7 & 72\% & 85\% & 90\% & 94\% & 97\% & 99\% & - & 99\% & 99\% \\ \cline{2-11} 
 & 8 & 73\% & 86\% & 90\% & 93\% & 96\% & 98\% & 99\% & - & 100\% \\ \cline{2-11} 
 & 9 & 74\% & 86\% & 90\% & 93\% & 98\% & 98\% & 99\% & 100\% & - \\ \hline
\end{tabular}
\end{table*}



 \subsection{Classification model generalized for multiple clients}

In the above sections, we considered different classification models for each web browser separately; one model for services accessed by Google Chrome and another one for services accessed by Firefox. However, for building a general classification model, we want to know if the browser used to train the model affects its validity for other sources of traffic. In theory, it should not because the features are designed to identify the service.

The authors in \cite{muehlstein2016analyzing} studied the effect of using different operating systems (Windows, Linux, Mac) and different web browsers (Internet Explorer, Google Chrome, Firefox) on the identification of some HTTPS services. Their results show that there is a strong relation between the client side environment (i.e., OS and Web browser) and the identification of the service of a given HTTPS flow (offline analysis). In our experiments, we use Ubuntu as operating system and, as mentioned earlier, Google Chrome and Firefox web browsers. Thus, the questions arises as to whether it is possible to train a classification model using HTTPS flows from Google Chrome to identify HTTPS flows from Firefox or vice-versa?

The experimental results show that, if we use Firefox's HTTPS flows for training the model and Google Chrome's HTTPS flows for testing, the identification accuracy is low (19\%) and this result is inline with \cite{muehlstein2016analyzing}. One possible reason of this low accuracy is that each browser uses a different implementation of TLS library: Mozilla uses Network Security Services (NSS) libraries to provide the support for TLS in Firefox web browser and other services \cite{networkmozilla}, while Google has developed its own fork of OpenSSL, named BoringSSL \cite{BoringSSL}. Using different implementation of the TLS protocol affects the collected features over different clients. This first observation tends to show that we need several separate models for identifying HTTPS services depending on the client. However, this approach complicates the early identification of HTTPS services, since we don't always have a prior knowledge about the used web browser. As a result, to provide a general classification model that is able to deal with HTTPS flows regardless the web browser, we have combined both Google Chrome and Firefox traces for training to generate a single and more generic model. For testing, we use this general model to assess if it can identify HTTPS services accessed from either Google Chrome or Firefox.

\begin{table*}[]
\setlength{\belowcaptionskip}{-5pt}
\footnotesize
\centering
\caption{Identification accuracy of a generic classification model\\ trained with both Google Chrome and Firefox flows}
\label{combinedmodel}
\begin{tabular}{|c|l|c|c|c|c|c|c|c|c|c|}
\hline
 & \multicolumn{1}{c|}{} & \multicolumn{9}{c|}{Testing (up to $d$) application data packets} \\ \cline{3-11} 
\multirow{-2}{*}{\begin{tabular}[c]{@{}c@{}}\#App. for \\ Training\end{tabular}} & \multicolumn{1}{c|}{\multirow{-2}{*}{\begin{tabular}[c]{@{}c@{}}{\scriptsize Web} \\ {\scriptsize Browser}\end{tabular}}} & \textbf{1} & \textbf{2} & \textbf{3} & \textbf{4} & 5 & \textbf{6} & \textbf{7} & \textbf{8} & \textbf{9} \\ \hline 
 & Chrome & - & 97\% & 95\% & 90\% & 89\% & 89\% & 88\% & 88\% & 88\% \\ \cline{2-11} 
\multirow{-2}{*}{up to 1} & Firefox & - & 93\% & 88\% & 86\% & 84\% & 84\% & 84\% & 83\% & 83\% \\ \hline \hline
 & Chrome & 82\% & - & 95\% & 93\% & 92\% & 92\% & 92\% & 91\% & 91\% \\ \cline{2-11} 
\multirow{-2}{*}{up to 2} & Firefox & 80\% & - & 94\% & 92\% & 91\% & 90\% & 89\% & 89\% & 89\% \\ \hline \hline
 & Chrome & 77\% & 93\% & - & 97\% & 95\% & 94\% & 93\% & 92\% & 91\% \\ \cline{2-11} 
\multirow{-2}{*}{up to 3} & Firefox & 69\% & 90\% & - & 96\% & 94\% & 93\% & 91\% & 90\% & 89\% \\ \hline \hline
 & Chrome & 73\% & 90\% & 95\% & - & 97\% & 97\% & 95\% & 93\% & 93\% \\ \cline{2-11} 
\multirow{-2}{*}{up to 4} & Firefox & 69\% & 87\% & 95\% & - & 97\% & 95\% & 93\% & 92\% & 91\% \\ \hline \hline
 & Chrome & 72\% & 87\% & 91\% & 97\% & - & 98\% & 97\% & 95\% & 95\% \\ \cline{2-11} 
\multirow{-2}{*}{up to 5} & Firefox & 68\% & 86\% & 92\% & 96\% & - & 98\% & 95\% & 95\% & 95\% \\ \hline 
\end{tabular}
\end{table*}

\begin{figure*} 
\setlength{\belowcaptionskip}{-8pt}
\centering
\mbox{
\subfloat[HTTPS services accessed by Google Chrome]{
\begin{tikzpicture}
\begin{axis}[ 	
	height=4.5cm, ylabel near ticks,
	width=6.5cm,
	tick label style={font=\scriptsize},
	label style={font=\scriptsize},
	xlabel= \# application data packets in the flow, 
	ylabel= Identification Accuracy (\%),style={font=\scriptsize},
	grid=major,xtick=data,
	ymax=100, 
	legend style={at={(0.95,0.3)}}
]
\addplot[color=red,mark=square] coordinates{
(1,75) (2,89) (3,94)(4,97)(6,98)(7,97)(8,95) (9,94)};
\addplot[color=blue,mark=triangle]  coordinates{
(1,72) (2,87) (3,91)(4,97)(6,98)(7,97)(8,95) (9,95)};
\legend{Special Chrome Model, Generic Model}
\end{axis}
\end{tikzpicture}
\label{chromevsgerneal}
}
\quad
\subfloat[HTTPS services accessed by Firefox]{
\begin{tikzpicture}
\begin{axis}[ 	
	height=4.5cm, ylabel near ticks,
	width=6.5cm,
	tick label style={font=\scriptsize},
	label style={font=\scriptsize},
	xlabel= \# application data packets in the flow,
	ylabel= Identification Accuracy (\%),style={font=\scriptsize},
	grid=major,xtick=data,
	ymax=100, 
	legend style={at={(0.95,0.3)}, font=\scriptsize}
]

\addplot[color=red,mark=otimes] coordinates{
(1,73)(2,88)(3,94)(4,97)(6,98)(7,97)(8,96)(9,96)};
\addplot[color=blue,mark=triangle]  coordinates{
(1,68)(2,86)(3,92)(4,96)(6,98)(7,95)(8,95)(9,95)};
\legend{Special Firefox Model, Generic Model}
\end{axis}
\end{tikzpicture}
\label{firfoxvsgerneal}
}
}
\caption{Dedicated classification models vs. the generic classification model}
\label{generalmodel}        
\end{figure*}
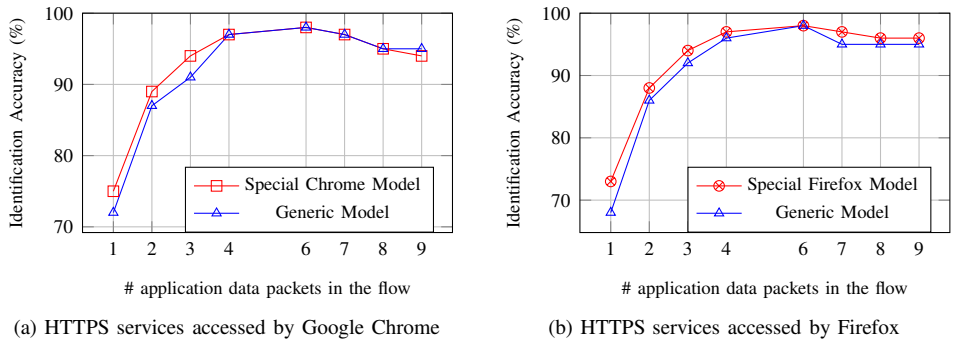

Table \ref{combinedmodel} shows the identification accuracy of using one general model trained with both Google Chrome and Firefox flows. For testing, we use flows from each web browser separately. We observe that the behaviour of the general classification model is similar to the dedicated ones. It means, with a low training threshold the accuracy decreases when testing threshold increases, but with a high training threshold, accuracy decreases even faster when the testing threshold lowers. So we have decided to consider using a training threshold of 5 application data packets to build a general model. 
Figures \ref{chromevsgerneal}, \ref{firfoxvsgerneal} compare the performance of the general model to identify HTTPS services accessed by Google Chrome and Firefox against a the dedicated models per web browser. We notice that the overall identification does not change significantly by using the general model. Based on these results, we conclude that it is possible to benefit from using one general model to early identify HTTPS services regardless of the web browser used to access a given service if the training dataset already represents clients' diversity.

\section{A prototype for early identification of services in HTTPS traffic}

\label{sec:realprotoype}
As a direct application of the previous results, we have developed a prototype that is able to recognize HTTPS services very early in the session. The prototype operates by extending the iptables/netfilter architecture. So it can be easily integrated into existing Linux middleboxes. 
Our software is released\footnote{https://gitlab.inria.fr/swazen/HTTPSFirewall} in open source for the community.

\subsection{Prototype architecture}
The proposed prototype operates in three levels; the first is hardware level, where packets are sniffed using iptables rules toward the kernel space (second level) and then the packets are copied in the user defined space (third level) for multiplexing. Hence, we divide the prototype into three modules; the packet capture module, the reassembly module, and the machine learning identification module. 


\subsubsection*{Packet capture module}
HTTPS packets are captured and forwarded from the iptables queue   to the user space (i.e., application level). We benefit from the libnetfilter\_queue\footnote{\url{https://github.com/chifflier/nfqueue-bindings}} to gain access to HTTPS packets queued using iptables rules, which forwards all input packets (from the remote HTTPS service) and output packets (from the local web browser) with port number 443 to the reassembly module to make a copy, and re-inject them back to the network level to allow the communication to continue. Then the arrived packets are handled by the reassembly module.

\subsubsection*{The reassembly module}
As explained before, we apply the proposed approach in \cite{ong2009} to optimize the HTTPS flow reassembly module. Taking into account that the monitoring system needs to handle thousands of TCP flows, we use the dictionary class to store the flows' record. The dictionary is accessed by a key that should be unique for each flow (i.e., the 4-tuples). Based on that, the TCP flow is presented as a list of TCP packet objects, which are stored based on their arrival order and can be accessed using the flow-key. Upon receiving TCP packets from the netfilter queue, the system calculates a hash key using the 4-tuple (source IP address, source port, destination IP address, destination port). This key is used to find the corresponding flow in the dictionary object. 



\subsubsection*{Machine learning identification module}
The identification module has been implemented using the scikit-learn\cite{scikit-learn}. The module's tasks are: to extract features (given in \Cref{sec:machine}) for a given HTTPS flow reconstructed by the reassembly module, and to run the pre-built model to predict the service.

\subsection{Performance evaluation}
The evaluation environment consists of a client machine configured to automatically open HTTPS websites, while our prototype runs on anther machine with 64-bit Intel Xeon Processor@3.80GHz, 32GB memory and OS Linux-Mint 17.0. 


First, we measure the amount of delay needed for identifying HTTPS services in real-time using TLS handshake packets and 9 application data packets (i.e., the worst case). This delay must be short, so we do not alter the browsing experience of users that access HTTPS websites through our identification framework. There are two potential sources of delay: the packet capture process using iptables and the machine learning identification procedure (i.e., the features extraction and the prediction phase). Concerning the first source of delay, as we just make a copy of captured packets without any treatment, we consider that this delay is negligible, and we only focus on the second one. The results show that we have in average 2.02$\pm0.86$ ms delay to process a single HTTPS flow, from the trigger of the flow reassembly to the final identification. According to \cite{naylor2014cost}, using HTTPS instead of HTTP adds an extra delay of 500 ms when using fibre links, while over 3G mobile network the delay is up to 1.2 seconds (depending on the RTT value). So the delay added by the framework is negligible for users because it is several orders of magnitude lower than the latency introduced by the protocol itself.

Second, to measure the identification accuracy of our approach we need to integrate the overhead delay of capturing packets and reconstructing flows, to make it sure that the training and testing will be done in the same conditions. Unfortunately, the public dataset was not captured through the framework and consequently this traffic cannot be used to evaluate the early identification prototype but only the model on which it is based (see \Cref{sec:apppacketsnum}). Indeed, TCP packets were dumped from the NIC to the dump file with no further delay. However in our prototype, TCP packets are captured from the NIC to netfilter queue (i.e., iptable) then forwarded to our framework for copy and re-injected back to the NIC. This process adds a significant delay to the Inter-packets arrival time which makes the features calculated for training and the one for testing incomparable because of this bias.

A traffic collector has been built to collect traffic in the same way of the identification prototype. 
For this new evaluation focused on the prototype implementation, a smaller dataset is used- top 100 HTTPS websites. The training was configured to use a threshold of 5 application data packets. The collected dataset contains 109513 records related to 582 HTTPS services. We use the cross-validation (k=10) and C4.5 to evaluate the identification accuracy. 

The results show that we have 95.96\% of accuracy. This higher accuracy compared to the evaluation framework in section \label{sec:evaluation} can be explained because the top 100 HTTPS services are easier to differentiate than the top 1000 used to evaluate the models: the more services, the higher the probability than a couple of them show a similar signature. Regarding the flow throughput, the proposed prototype is able to process at a rate of 31874 flow/sec. This throughput is higher than the flow throughput given in \cite{PEREIRA2014} (24997 flow/sec) because we need to use a smaller number of packets per flow before starting the identification process. Based on these results, the prototype performs well with a low delay overhead and a high flow throughput, and also delivers a good identification accuracy. 


\section{Conclusion}
\label{sec:conclusion}
The increasing usage of HTTPS undermines the effectiveness of service-level monitoring approaches for a large part of Internet traffic. In this paper, we tackled the early identification of HTTPS services. We proposed a new machine learning-based method which is able to identify HTTPS services with a high accuracy by using statistical features on the TLS handshake packets and a small number of application data packets. The proposed approach was evaluated using our public HTTPS dataset-built to solve the issue of a reference HTTPS dataset.
Extensive evaluation results show that we can identify HTTPS services using the TLS handshake packets and up to 9 application data packets for long flows, but 5 packets is a better tradeoff to identify both short and long flows. As a direct application of our results, we presented an identification prototype to identify HTTPS service. Experimental tests show that our approach has  alow overhead delay of 2.02 msec and high flow throughput 31874 fps. In future work, we will adapt our approach to identify services over HTTP2.0.

\bibliographystyle{IEEEtran}
\bibliography{references}

\begin{thebibliography}{10}
\providecommand{\url}[1]{#1}
\csname url@samestyle\endcsname
\providecommand{\newblock}{\relax}
\providecommand{\bibinfo}[2]{#2}
\providecommand{\BIBentrySTDinterwordspacing}{\spaceskip=0pt\relax}
\providecommand{\BIBentryALTinterwordstretchfactor}{4}
\providecommand{\BIBentryALTinterwordspacing}{\spaceskip=\fontdimen2\font plus
\BIBentryALTinterwordstretchfactor\fontdimen3\font minus
  \fontdimen4\font\relax}
\providecommand{\BIBforeignlanguage}[2]{{%
\expandafter\ifx\csname l@#1\endcsname\relax
\typeout{** WARNING: IEEEtran.bst: No hyphenation pattern has been}%
\typeout{** loaded for the language `#1'. Using the pattern for}%
\typeout{** the default language instead.}%
\else
\language=\csname l@#1\endcsname
\fi
#2}}
\providecommand{\BIBdecl}{\relax}
\BIBdecl

\bibitem{McCarthy2011}
C.~McCarthy \emph{et~al.}, ``An investigation on identifying {SSL} traffic,''
  in \emph{Computational Intelligence for Security and Defense Applications
  (CISDA), 2011 IEEE Symposium on}.\hskip 1em plus 0.5em minus 0.4em\relax
  IEEE, 2011, pp. 115--122.

\bibitem{aertsen2016no}
M.~Aertsen, M.~Korczy{\'n}ski, G.~Moura, S.~Tajalizadehkhoob, and J.~v.~d.
  Berg, ``No domain left behind: is let's encrypt democratizing encryption?''
  \emph{arXiv preprint arXiv:1612.03005}, 2016.

\bibitem{cisco2016}
``{Cisco 2016 Annual Security Report},''
  \url{http://www.cisco.com/c/dam/assets/offers/pdfs/cisco-asr-2016.pdf},
  (Online, {retrieved}: 30 May, 2016).

\bibitem{Avis2015}
``Report 2015-0832 from the {F}rench regulatory authority for
  telecommunications ({ARCEP}). {S}tructure de l'usage de la bande passante des
  reseaux d'acces a internet sur le territoire francais,''
  "\url{http://www.arcep.fr/uploads/tx_gsavis/15-0832.pdf}", ({[In French]},
  {retrieved}: 07 July, 2015).

\bibitem{Sandvine2016}
``Global internet phenomena spotlight: Encrypted internet traffic,''
  \url{https://www.sandvine.com/downloads/general/global-internet-phenomena/2015/encrypted-internet-traffic.pdf},
  (Online, {retrieved}: 11 May, 2016).

\bibitem{mozilla2016}
``{HTTPS} adoption has reached the tipping point,''
  \url{https://www.troyhunt.com/https-adoption-has-reached-the-tipping-point/},
  (Online, {retrieved}: 3 February, 2017).

\bibitem{GoogleReport2016}
``{Google Transparency Report, {HTTPS} Usage},''
  \url{https://www.google.com/transparencyreport/https/metrics/?hl=en},
  (Online, {retrieved}: 18 December, 2016).

\bibitem{netcraft2014}
``January 2014 web server survey,''
  "\url{http://news.netcraft.com/archives/2014/01/03/january-2014-web-server-survey.html}",
  (Online, {retrieved}: 29 August, 2016).

\bibitem{mori2016statistical}
T.~Mori, T.~Inoue, A.~Shimoda, K.~Sato, S.~Harada, K.~Ishibashi, and S.~Goto,
  ``Statistical estimation of the names of https servers with domain name
  graphs,'' \emph{Computer Communications}, 2016.

\bibitem{Google2016}
``{HTTPS} as a ranking signal,''
  \url{https://webmasters.googleblog.com/2014/08/https-as-ranking-signal.html},
  (Online, {retrieved}: 3 September, 2016).

\bibitem{Kaspersky2014}
``{HTTPS} working for malicious users,''
  \url{https://securelist.com/blog/research/57323/https-working-for-malicious-users},
  (Online, {retrieved}: 3 September, 2016).

\bibitem{chen2014security}
P.~Chen, N.~Nikiforakis, L.~Desmet, and C.~Huygens, ``Security analysis of the
  chinese web: How well is it protected?'' in \emph{Proceedings of the 2014
  Workshop on Cyber Security Analytics, Intelligence and Automation}.\hskip 1em
  plus 0.5em minus 0.4em\relax ACM, 2014, pp. 3--9.

\bibitem{pukkawanna2014classification}
S.~Pukkawanna, G.~Blanc, J.~Garcia-Alfaro, Y.~Kadobayashi, and H.~Debar,
  ``Classification of {SSL} servers based on their {SSL} handshake for
  automated security assessment,'' in \emph{2014 Third International Workshop
  on Building Analysis Datasets and Gathering Experience Returns for Security
  (BADGERS)}.\hskip 1em plus 0.5em minus 0.4em\relax IEEE, 2014, pp. 30--39.

\bibitem{bortolameotti2015indicators}
R.~Bortolameotti, A.~Peter, M.~H. Everts, and D.~Bolzoni, ``Indicators of
  malicious {SSL} connections,'' in \emph{Network and System Security}.\hskip
  1em plus 0.5em minus 0.4em\relax Springer, 2015, pp. 162--175.

\bibitem{shbair2015efficiently}
W.~Shbair, T.~Cholez, A.~Goichot, and I.~Chrisment, ``Efficiently bypassing
  {SNI}-based {HTTPS} filtering,'' in \emph{IFIP/IEEE International Symposium
  on Integrated Network Management (IM)}.\hskip 1em plus 0.5em minus
  0.4em\relax IEEE, 2015, pp. 990--995.

\bibitem{dierks2008transport}
T.~Dierks and E.~Rescorla, ``{RFC} 5246 - the transport layer security ({TLS})
  protocol version 1.2,'' \url{http://tools.ietf.org/html/rfc5246}, 2008.

\bibitem{ong2009}
B.~Xiong, C.~Xiao-su, and C.~Ning, ``A real-time {TCP} stream reassembly
  mechanism in high-speed network,'' \emph{South West Jiaotong University},
  vol.~17, no.~3, pp. 185--191, 2009.

\bibitem{PEREIRA2014}
S.~S.~L. Pereira, J.~L. d.~C. e~Silva, and J.~E.~B. Maia, ``{NTCS}: A real time
  flow-based network traffic classification system,'' in \emph{10th
  International Conference on Network and Service Management (CNSM) and
  Workshop}.\hskip 1em plus 0.5em minus 0.4em\relax IEEE, 2014, pp. 368--371.

\bibitem{groleat2014high}
T.~Grol{\'e}at, S.~Vaton, and M.~Arzel, ``High-speed flow-based classification
  on {FPGA},'' \emph{International journal of network management}, vol.~24,
  no.~4, pp. 253--271, 2014.

\bibitem{bernaille2007early}
L.~Bernaille and R.~Teixeira, ``Early recognition of encrypted applications,''
  in \emph{Passive and Active Network Measurement}.\hskip 1em plus 0.5em minus
  0.4em\relax Springer, 2007, pp. 165--175.

\bibitem{kumano2014towards}
Y.~Kumano, S.~Ata, N.~Nakamura, Y.~Nakahira, and I.~Oka, ``Towards real-time
  processing for application identification of encrypted traffic,'' in
  \emph{Computing, Networking and Communications (ICNC), 2014 International
  Conference on}.\hskip 1em plus 0.5em minus 0.4em\relax IEEE, 2014, pp.
  136--140.

\bibitem{MAIOLINI2009}
G.~Maiolini, A.~Baiocchi, A.~Iacovazzi, and A.~Rizzi, ``Real time
  identification of {SSH} encrypted application flows by using cluster analysis
  techniques,'' in \emph{International Conference on Research in
  Networking}.\hskip 1em plus 0.5em minus 0.4em\relax Springer, 2009, pp.
  182--194.

\bibitem{Yanai2010}
R.~Bar-Yanai, M.~Langberg, D.~Peleg, and L.~Roditty, ``Realtime classification
  for encrypted traffic,'' in \emph{International Symposium on Experimental
  Algorithms}.\hskip 1em plus 0.5em minus 0.4em\relax Springer, 2010, pp.
  373--385.

\bibitem{nguyen2008survey}
T.~T. Nguyen and G.~Armitage, ``A survey of techniques for internet traffic
  classification using machine learning,'' \emph{Communications Surveys \&
  Tutorials, IEEE}, vol.~10, no.~4, pp. 56--76, 2008.

\bibitem{Jun2008}
J.~Li, S.~Zhang, Y.~Lu, and J.~Yan, ``Real-time {P2P} traffic identification,''
  in \emph{IEEE GLOBECOM 2008-2008 IEEE Global Telecommunications
  Conference}.\hskip 1em plus 0.5em minus 0.4em\relax IEEE, 2008, pp. 1--5.

\bibitem{Chengjie2011}
C.~Gu, S.~Zhang, and Y.~Sun, ``Realtime encrypted traffic identification using
  machine learning,'' \emph{Journal of Software}, vol.~6, no.~6, pp.
  1009--1016, 2011.

\bibitem{Bernaille2006}
L.~Bernaille, R.~Teixeira, I.~Akodkenou, A.~Soule, and K.~Salamatian, ``Traffic
  classification on the fly,'' \emph{ACM SIGCOMM Computer Communication
  Review}, vol.~36, no.~2, pp. 23--26, 2006.

\bibitem{xue_traffic_2013}
Y.~Xue, D.~Wang, and L.~Zhang, ``Traffic classification: Issues and
  challenges,'' in \emph{2013 International Conference on Computing, Networking
  and Communications ({ICNC)}}, 2013, pp. 545--549.

\bibitem{alshammari2009machine}
R.~Alshammari \emph{et~al.}, ``Machine learning based encrypted traffic
  classification: identifying {SSH} and {Skype},'' in \emph{Computational
  Intelligence for Security and Defense Applications, 2009. CISDA 2009. IEEE
  Symposium on}.\hskip 1em plus 0.5em minus 0.4em\relax IEEE, 2009, pp. 1--8.

\bibitem{mccarthy_investigation_2011}
C.~McCarthy \emph{et~al.}, ``An investigation on identifying {SSL} traffic,''
  in \emph{Computational Intelligence for Security and Defense Applications
  (CISDA), 2011 IEEE Symposium on}.\hskip 1em plus 0.5em minus 0.4em\relax
  IEEE, 2011, pp. 115--122.

\bibitem{li2007identifying}
J.~Li, S.~Zhang, Y.~Xuan, and Y.~Sun, ``Identifying {Skype} traffic by random
  forest,'' in \emph{2007 International Conference on Wireless Communications,
  Networking and Mobile Computing}.\hskip 1em plus 0.5em minus 0.4em\relax
  IEEE, 2007, pp. 2841--2844.

\bibitem{singh2013near}
K.~Singh, S.~Agrawal, and B.~Sohi, ``A near real-time {IP} traffic
  classification using machine learning,'' \emph{International Journal of
  Intelligent Systems and Applications}, vol.~5, no.~3, p.~83, 2013.

\bibitem{Williams2006}
N.~Williams, S.~Zander, and G.~Armitage, ``A preliminary performance comparison
  of five machine learning algorithms for practical {IP} traffic flow
  classification,'' \emph{ACM SIGCOMM Computer Communication Review}, vol.~36,
  no.~5, pp. 5--16, 2006.

\bibitem{gu2011realtime}
C.~Gu, S.~Zhang, and Y.~Sun, ``Realtime encrypted traffic identification using
  machine learning,'' \emph{Journal of Software}, vol.~6, no.~6, pp.
  1009--1016, 2011.

\bibitem{hullar2011early}
B.~Hull{\'a}r, S.~Laki, and A.~Gyorgy, ``Early identification of peer-to-peer
  traffic,'' in \emph{2011 IEEE International Conference on Communications
  (ICC)}.\hskip 1em plus 0.5em minus 0.4em\relax IEEE, 2011, pp. 1--6.

\bibitem{Velan2015}
P.~Velan, M.~{\v{C}}erm{\'a}k, P.~{\v{C}}eleda, and M.~Dra{\v{s}}ar, ``A survey
  of methods for encrypted traffic classification and analysis,''
  \emph{International Journal of Network Management}, vol.~25, no.~5, pp.
  355--374, 2015.

\bibitem{qian2012flow}
L.~Qian and B.~E. Carpenter, ``A flow-based performance analysis of {TCP} and
  {TCP} applications,'' in \emph{2012 18th IEEE International Conference on
  Networks (ICON)}.\hskip 1em plus 0.5em minus 0.4em\relax IEEE, 2012, pp.
  41--45.

\bibitem{muehlstein2016analyzing}
J.~Muehlstein, Y.~Zion, M.~Bahumi, I.~Kirshenboim, R.~Dubin, A.~Dvir, and
  O.~Pele, ``Analyzing {HTTPS} encrypted traffic to identify user operating
  system, browser and application,'' \emph{arXiv preprint arXiv:1603.04865},
  2016.

\bibitem{networkmozilla}
M.~D. Network, ``Mozilla network security services (nss),''
  \url{https://developer.mozilla.org/en-US/docs/Mozilla/Projects/NSS}.

\bibitem{BoringSSL}
``Boringssl,'' \url{https://boringssl.googlesource.com/boringssl/}, (Online,
  {retrieved}: 13 December, 2016).

\bibitem{Naylor2014}
D.~Naylor, A.~Finamore, I.~Leontiadis, Y.~Grunenberger, M.~Mellia,
  M.~Munaf{\`o}, K.~Papagiannaki, and P.~Steenkiste, ``The cost of the {S} in
  {HTTPS},'' in \emph{Proceedings of the 10th ACM International on Conference
  on emerging Networking Experiments and Technologies}.\hskip 1em plus 0.5em
  minus 0.4em\relax ACM, 2014, pp. 133--140.

\bibitem{schatzmann2010digging}
D.~Schatzmann, W.~M{\"u}hlbauer, T.~Spyropoulos, and X.~Dimitropoulos,
  ``Digging into {HTTPS}: flow-based classification of webmail traffic,'' in
  \emph{Proceedings of the 10th ACM SIGCOMM conference on Internet
  measurement}.\hskip 1em plus 0.5em minus 0.4em\relax ACM, 2010, pp. 322--327.

\end{thebibliography}


\begin{thebibliography}{10}
\providecommand{\url}[1]{#1}
\csname url@samestyle\endcsname
\providecommand{\newblock}{\relax}
\providecommand{\bibinfo}[2]{#2}
\providecommand{\BIBentrySTDinterwordspacing}{\spaceskip=0pt\relax}
\providecommand{\BIBentryALTinterwordstretchfactor}{4}
\providecommand{\BIBentryALTinterwordspacing}{\spaceskip=\fontdimen2\font plus
\BIBentryALTinterwordstretchfactor\fontdimen3\font minus
  \fontdimen4\font\relax}
\providecommand{\BIBforeignlanguage}[2]{{%
\expandafter\ifx\csname l@#1\endcsname\relax
\typeout{** WARNING: IEEEtran.bst: No hyphenation pattern has been}%
\typeout{** loaded for the language `#1'. Using the pattern for}%
\typeout{** the default language instead.}%
\else
\language=\csname l@#1\endcsname
\fi
#2}}
\providecommand{\BIBdecl}{\relax}
\BIBdecl

\bibitem{McCarthy2011}
C.~McCarthy \emph{et~al.}, ``An investigation on identifying {SSL} traffic,''
  in \emph{Computational Intelligence for Security and Defense Applications
  (CISDA), 2011 IEEE Symposium on}.\hskip 1em plus 0.5em minus 0.4em\relax
  IEEE, 2011, pp. 115--122.

\bibitem{Cisco2018}
``{Cisco 2018 Annual Cypersecurity Report},''
  \url{https://www.cisco.com/c/dam/m/hu_hu/campaigns/security-hub/pdf/acr-2018.pdf},
  (Online, {retrieved}: 13 July, 2018).

\bibitem{mozilla2016}
``{HTTPS} adoption has reached the tipping point,''
  \url{https://www.troyhunt.com/https-adoption-has-reached-the-tipping-point/},
  (Online, {retrieved}: 3 February, 2017).

\bibitem{Kaspersky2014}
``{HTTPS} working for malicious users,''
  \url{https://securelist.com/blog/research/57323/https-working-for-malicious-users},
  (Online, {retrieved}: 3 September, 2016).

\bibitem{chen2014security}
P.~Chen, N.~Nikiforakis, L.~Desmet, and C.~Huygens, ``Security analysis of the
  chinese web: How well is it protected?'' in \emph{Proceedings of the 2014
  Workshop on Cyber Security Analytics, Intelligence and Automation}.\hskip 1em
  plus 0.5em minus 0.4em\relax ACM, 2014, pp. 3--9.

\bibitem{pukkawanna2014classification}
S.~Pukkawanna, G.~Blanc, J.~Garcia-Alfaro, Y.~Kadobayashi, and H.~Debar,
  ``Classification of {SSL} servers based on their {SSL} handshake for
  automated security assessment,'' in \emph{2014 Third International Workshop
  on Building Analysis Datasets and Gathering Experience Returns for Security
  (BADGERS)}.\hskip 1em plus 0.5em minus 0.4em\relax IEEE, 2014, pp. 30--39.

\bibitem{shbair2015efficiently}
W.~Shbair, T.~Cholez, A.~Goichot, and I.~Chrisment, ``Efficiently bypassing
  {SNI}-based {HTTPS} filtering,'' in \emph{IFIP/IEEE International Symposium
  on Integrated Network Management (IM)}.\hskip 1em plus 0.5em minus
  0.4em\relax IEEE, 2015, pp. 990--995.

\bibitem{nguyen2008survey}
T.~T. Nguyen and G.~Armitage, ``A survey of techniques for internet traffic
  classification using machine learning,'' \emph{Communications Surveys \&
  Tutorials, IEEE}, vol.~10, no.~4, pp. 56--76, 2008.

\bibitem{shbair2016multi}
W.~M. Shbair, T.~Cholez, J.~Francois, and I.~Chrisment, ``A multi-level
  framework to identify {HTTPS} services,'' in \emph{Network Operations and
  Management Symposium (NOMS), 2016 IEEE/IFIP}, pp. 240--248.

\bibitem{ong2009}
B.~Xiong, C.~Xiao-su, and C.~Ning, ``A real-time {TCP} stream reassembly
  mechanism in high-speed network,'' \emph{South West Jiaotong University},
  vol.~17, no.~3, pp. 185--191, 2009.

\bibitem{PEREIRA2014}
S.~S.~L. Pereira, J.~L. d.~C. e~Silva, and J.~E.~B. Maia, ``{NTCS}: A real time
  flow-based network traffic classification system,'' in \emph{10th
  International Conference on Network and Service Management (CNSM) and
  Workshop}.\hskip 1em plus 0.5em minus 0.4em\relax IEEE, 2014, pp. 368--371.

\bibitem{groleat2014high}
T.~Grol{\'e}at, S.~Vaton, and M.~Arzel, ``High-speed flow-based classification
  on {FPGA},'' \emph{International journal of network management}, vol.~24,
  no.~4, pp. 253--271, 2014.

\bibitem{bernaille2007early}
L.~Bernaille and R.~Teixeira, ``Early recognition of encrypted applications,''
  in \emph{Passive and Active Network Measurement}.\hskip 1em plus 0.5em minus
  0.4em\relax Springer, 2007, pp. 165--175.

\bibitem{kumano2014towards}
Y.~Kumano, S.~Ata, N.~Nakamura, Y.~Nakahira, and I.~Oka, ``Towards real-time
  processing for application identification of encrypted traffic,'' in
  \emph{Computing, Networking and Communications (ICNC), International
  Conference on}.\hskip 1em plus 0.5em minus 0.4em\relax IEEE, 2014, pp.
  136--140.

\bibitem{MAIOLINI2009}
G.~Maiolini, A.~Baiocchi, A.~Iacovazzi, and A.~Rizzi, ``Real time
  identification of {SSH} encrypted application flows by using cluster analysis
  techniques,'' in \emph{International Conference on Research in
  Networking}.\hskip 1em plus 0.5em minus 0.4em\relax Springer, 2009, pp.
  182--194.

\bibitem{Yanai2010}
R.~Bar-Yanai, M.~Langberg, D.~Peleg, and L.~Roditty, ``Realtime classification
  for encrypted traffic,'' in \emph{International Symposium on Experimental
  Algorithms}.\hskip 1em plus 0.5em minus 0.4em\relax Springer, 2010, pp.
  373--385.

\bibitem{Jun2008}
J.~Li, S.~Zhang, Y.~Lu, and J.~Yan, ``Real-time {P2P} traffic identification,''
  in \emph{IEEE GLOBECOM 2008-2008 IEEE Global Telecommunications
  Conference}.\hskip 1em plus 0.5em minus 0.4em\relax IEEE, 2008, pp. 1--5.

\bibitem{Chengjie2011}
C.~Gu, S.~Zhang, and Y.~Sun, ``Realtime encrypted traffic identification using
  machine learning,'' \emph{Journal of Software}, vol.~6, no.~6, pp.
  1009--1016, 2011.

\bibitem{Bernaille2006}
L.~Bernaille, R.~Teixeira, I.~Akodkenou, A.~Soule, and K.~Salamatian, ``Traffic
  classification on the fly,'' \emph{ACM SIGCOMM Computer Communication
  Review}, vol.~36, no.~2, pp. 23--26, 2006.

\bibitem{xue_traffic_2013}
Y.~Xue, D.~Wang, and L.~Zhang, ``Traffic classification: Issues and
  challenges,'' in \emph{2013 International Conference on Computing, Networking
  and Communications ({ICNC)}}, 2013, pp. 545--549.

\bibitem{alshammari2009machine}
R.~Alshammari \emph{et~al.}, ``Machine learning based encrypted traffic
  classification: identifying {SSH} and {Skype},'' in \emph{Computational
  Intelligence for Security and Defense Applications, CISDA 2009. IEEE
  Symposium on}, pp. 1--8.

\bibitem{mccarthy_investigation_2011}
C.~McCarthy \emph{et~al.}, ``An investigation on identifying {SSL} traffic,''
  in \emph{Computational Intelligence for Security and Defense Applications
  (CISDA), 2011 IEEE Symposium on}.\hskip 1em plus 0.5em minus 0.4em\relax
  IEEE, 2011, pp. 115--122.

\bibitem{li2007identifying}
J.~Li, S.~Zhang, Y.~Xuan, and Y.~Sun, ``Identifying {Skype} traffic by random
  forest,'' in \emph{International Conference on Wireless Communications,
  Networking and Mobile Computing}.\hskip 1em plus 0.5em minus 0.4em\relax
  IEEE, 2007, pp. 2841--2844.

\bibitem{singh2013near}
K.~Singh, S.~Agrawal, and B.~Sohi, ``A near real-time {IP} traffic
  classification using machine learning,'' \emph{International Journal of
  Intelligent Systems and Applications}, vol.~5, no.~3, p.~83, 2013.

\bibitem{Williams2006}
N.~Williams, S.~Zander, and G.~Armitage, ``A preliminary performance comparison
  of five machine learning algorithms for practical {IP} traffic flow
  classification,'' \emph{ACM SIGCOMM Computer Communication Review}, vol.~36,
  no.~5, pp. 5--16, 2006.

\bibitem{Velan2015}
P.~Velan, M.~{\v{C}}erm{\'a}k, P.~{\v{C}}eleda, and M.~Dra{\v{s}}ar, ``A survey
  of methods for encrypted traffic classification and analysis,''
  \emph{International Journal of Network Management}, vol.~25, no.~5, pp.
  355--374, 2015.

\bibitem{qian2012flow}
L.~Qian and B.~E. Carpenter, ``A flow-based performance analysis of {TCP} and
  {TCP} applications,'' in \emph{2012 18th IEEE International Conference on
  Networks (ICON)}.\hskip 1em plus 0.5em minus 0.4em\relax IEEE, 2012, pp.
  41--45.

\bibitem{muehlstein2016analyzing}
J.~Muehlstein, Y.~Zion, M.~Bahumi, I.~Kirshenboim, R.~Dubin, A.~Dvir, and
  O.~Pele, ``Analyzing {HTTPS} encrypted traffic to identify user operating
  system, browser and application,'' \emph{arXiv}, 2016.

\bibitem{networkmozilla}
M.~D. Network, ``Mozilla network security services (nss),''
  \url{https://developer.mozilla.org/en-US/docs/Mozilla/Projects/NSS}.

\bibitem{BoringSSL}
``Boringssl,'' \url{https://boringssl.googlesource.com/boringssl/}, (Online,
  {retrieved}: 13 December, 2016).

\bibitem{scikit-learn}
F.~Pedregosa, G.~Varoquaux, A.~Gramfort, V.~Michel, B.~Thirion, O.~Grisel,
  M.~Blondel, P.~Prettenhofer, R.~Weiss, V.~Dubourg, J.~Vanderplas, A.~Passos,
  D.~Cournapeau, M.~Brucher, M.~Perrot, and E.~Duchesnay, ``Scikit-learn:
  Machine learning in {P}ython,'' \emph{Journal of Machine Learning Research},
  vol.~12, pp. 2825--2830, 2011.

\bibitem{naylor2014cost}
D.~Naylor, A.~Finamore, I.~Leontiadis, Y.~Grunenberger, M.~Mellia,
  M.~Munaf{\`o}, K.~Papagiannaki, and P.~Steenkiste, ``The cost of the s in
  https,'' in \emph{Proceedings of the 10th ACM International on Conference on
  emerging Networking Experiments and Technologies}.\hskip 1em plus 0.5em minus
  0.4em\relax ACM, 2014, pp. 133--140.

\end{thebibliography}

\end{document}